\documentclass[aps,prd,preprint,nofootinbib]{revtex4}
\usepackage{amsfonts}
\usepackage{mathrsfs}
\usepackage{graphicx}
\usepackage{amsmath}
\usepackage{amssymb}
\usepackage{subfigure}
\usepackage{epsfig}
\usepackage{graphicx}
\usepackage{slashed}
\usepackage{color}
\newcommand{\bqa}{\begin{eqnarray}}
\newcommand{\eqa}{\end{eqnarray}}
\newcommand{\beq}{\begin{equation}}
\newcommand{\eeq}{\end{equation}}
\allowdisplaybreaks[1]
\graphicspath{{fig/}{dia/}} \DeclareGraphicsExtensions{.eps}
\begin{document}

\title{The solution to the `$1/2$ vs $3/2$' puzzle}

\author{Guo-Li Wang$^{1,2}$\footnote[1]{wgl@hbu.edu.cn, corresponding author}, Qiang Li$^{3}$\footnote[1]{corresponding author}, Tianhong Wang$^{4}$\footnote[1]{corresponding author}, Tai-Fu Feng$^{1,2}$, Xing-Gang Wu$^{5}$, Chao-Hsi Chang$^{6,7}$}

\affiliation{$^1$ Department of Physics, Hebei University, Baoding 071002, China\\
$^2$ Key Laboratory of High-precision Computation and Application of Quantum Field Theory of Hebei Province, Baoding 071002, China\\
$^3$ School of Physical Science and Technology, Northwestern Polytechnical University, Xi'an 710072, China\\
$^4$ School of Physics, Harbin Institute of Technology, Harbin 150001, China\\
$^5$ Department of Physics, Chongqing Key Laboratory for Strongly Coupled Physics, Chongqing University, Chongqing 401331, China\\
$^6$ Institute of Theoretical Physics, Chinese Academy of Science, Beijing 100190, China\\
$^7$ CCAST(World Laboratory), P.O. Box 8730, Beijing 100190, China}

\begin{abstract}
\vspace{0.5cm}
Using an almost complete relativistic method based on the Bethe-Salpeter equation, we study the mixing angle $\theta$, the mass splitting $\bigtriangleup M$, the strong decay widths $\Gamma(D^{({\prime})}_1)$ and the weak production rates $Br(B\to D^{({\prime})}_1\ell\nu_{\ell})$ of the $D_1(2420)$ and $D_1^{\prime}(2430)$. We find there is the strong cancellation between the $^1P_1$ and $^3P_1$ partial waves in $D_1^{\prime}(2430)$ with $\theta \sim-35.3^{\circ}$, which leads to the `$1/2$ vs $3/2$' puzzle. The puzzle can not be overcome by adding only relativistic corrections since in a large parameter range where $\bigtriangleup M$ is linear varying and not small, the $\theta$, $\Gamma(D^{({\prime})}_1)$ and $Br(B\to D^{({\prime})}_1\ell\nu_{\ell})$ remain almost unchanged but conflict with data. While in a special range around the mass inverse point where $\bigtriangleup M=0$ and $\theta =\pm 90^{\circ}$, they change rapidly but we find the windows where $\bigtriangleup M$, $\Gamma(D^{({\prime})}_1)$ and $Br(B\to D^{({\prime})}_1\ell\nu_{\ell})$ are all consistent with data. The small $\bigtriangleup M$ confirmed by experiment, is crucial to solve the `$1/2$ vs $3/2$' puzzle.

\end{abstract}
 \maketitle
\newpage

\section{Introduction}\label{Sec-1}
In the $B$ semi-leptonic decays, there is a long-lived puzzle, which is the `$1/2$ vs $3/2$' puzzle \cite{puzzle,bigi,scora,colangelo,huang,ebert1}.
That is, theoretical calculations predict that semi-leptonic $B$ decays should have a substantially smaller rate to the ${\frac{1}{2}}^+$ than to the ${\frac{3}{2}}^+$ doublet,
$Br(B\to {D}^{{1}/{2}}_{0,1}\ell\nu_{\ell})\ll Br(B\to {D}^{{3}/{2}}_{1,2}\ell\nu_{\ell})$, which conflicts with experimental results, $Br(B\to {D}^{{1}/{2}}_{0,1}\ell\nu_{\ell})\approx Br(B\to {D}^{{3}/{2}}_{1,2}\ell\nu_{\ell})$ and inspires a lot of research interests \cite{Leibovich,Leibovich2,vito,beci,bernlochner2,benoit}. Many efforts have been made to overcome this difficulty \cite{Bernlochner,liqiang2,blossier,klein,yaouan}, and part of the theoretical-experimental differences, for example, the $B\to {D}_{0}\ell\nu_{\ell}$ decay could be improved by adding relativistic corrections \cite{ebert,jiangyue}, while the puzzle regarding the two $1^+$ states $D_1$ and $D_1^{\prime}$ remains to this day.

Since there is a light quark in $B$ and $P$ wave $D_J$ ($J=0,1,2$) mesons, the relativistic corrections are expected large in the transition $B\to D_J$.
Actually, we have shown that even in a double heavy $0^- \to 0^+$ process $B_c\to {\chi}_{c0}\ell\nu_{\ell}$, the relativistic corrections are very large \cite{gengzikan,gengzikan2}. Our previous study confirms that the $Br(B\to {D}_{0}\ell\nu_{\ell})$ can be enhanced substantially \cite{jiangyue} and consist with data. For the process $B\to {D}_{2}\ell\nu_{\ell}$, the relativistic corrections almost cancel each other \cite{gengzikan}. So in a relativistic study, the $Br(B\to {D}_{2}\ell\nu_{\ell})$ agrees with data \cite{jiangyue}. While for the two $1^+$ final states, choosing our usually used parameters, we find large cancellation between $^1P_1$ and $^3P_1$ partial waves in $D^{\prime}_1$ state, which leads to an one order smaller $Br(B\to D^{\prime}_1 \ell\nu_{\ell})$ than $Br(B\to {D}_{1} \ell\nu_{\ell})$, see the figure \ref{weak}. Other relativistic studies also obtained the similar results \cite{ebert,dong}. So relativistic corrections, or similarly, $1/m$ corrections in Heavy Quark Effective Theory (HQET) \cite{blossier} can not overcome the `$1/2$ vs $3/2$' puzzle.

In addition, when calculate the strong decays, we find the $D_1^{\prime}(2430)$ strong decay width is within the range of data, while the $D_1(2420)$ width is about half of the experimental data, see figure \ref{strong}. So both the strong decay widths and the weak production rates in theory do not agree with data. We know that $1^+$ states $D_1$ and $D^{\prime}_1$ are mixture of $^1P_1$ and $^3P_1$, and Ref. \cite{klein} pointed out that the mixing may soften the `$1/2$ vs $3/2$' puzzle, therefore,
we wonder whether this puzzle could be solved by using a complete relativistic method with the help of such mixing property.

In previous paper \cite{liqiang}, we solved the complete Salpeter equation for a $1^+$ state, and obtained almost complete relativistic wave function (except the instantaneous approximation), which provides us a new way to calculate the mixing angle. We find this almost complete relativistic method brings new phenomena, namely the obtained meson masses inverse and the the mixing angle flips in some parameter spaces. At the same time, it also leads to great changes of some physical quantities in this range. So we wonder whether this new phenomenon could help to explain the `$1/2$ vs $3/2$' puzzle.

In Sect. II, we will give the relativistic wave functions used and our method to calculate the mixing angle for $1^+$ states. The methods to calculate the transition matrix elements for weak and strong decays are shown in Sect. III. Our results are given in Sect. IV. Finally, we give a short summary in Sect. V.
\section{Relativistic wave functions}
In this section, we give the relativistic wave functions of the bound states used in this paper. The quantum number $J^P$ of each term in the wave function is the same as that of the meson. The relativistic wave function of a meson is a function of momentum $P$, mass $M$, possible polarization $\varepsilon$ of the meson, as well as the relative momentum $q$ between quark 1 and anti-quark 2 with ${q}_{_{\bot}}=q-\frac{q\cdot P}{M^2}P$ ($=(0,\vec{q})$ in the center-of-mass frame of the meson), the quark mass $m_1$, anti-quark mass $m_2$, quark energy $\omega_1$, and anti-quark energy $\omega_2$ with $\omega_i=\sqrt{m_i^2+\vec q\,^2}$ ($i=1,2$).
\subsection{Wave function of a $1^+$ state and its mixing angle}
In a previous paper \cite{liqiang}, by solving the complete instantaneous Bethe-Salpeter equation \cite{bs,sal} for a $1^+$ state, we obtain the relativistic wave function for $D_1$ or $D^{\prime}_1$,
\begin{equation}
\begin{aligned}\label{wave}
\varphi^{1^{+}}_{_P}({q}_{_{\bot}})=\displaystyle \varepsilon\cdot {q}_{_{\bot}}\left(g^{}_1+g^{}_2\frac{\not\!P}{M}-{g_1 x_{-}\not\!{q}_{_{\bot}}}
+\frac{g^{}_2 x_{+}\not\!{q}_{_{\bot}}\not\!P}{{M}}\right)\gamma^{5}
\\
+\displaystyle\frac{i}{M}\left(h^{}_1+h^{}_2\frac{\not\!P}{M}-{h^{}_1 x_{-}\not\!{q}_{_{\bot}}}
+\frac{h^{}_2 x_{+}\not\!{q}_{_{\bot}}\not\!P}{M}\right)\epsilon^{}_{\nu\lambda\rho\sigma}\gamma^{\nu}_{}
P^{\lambda}_{}q^{\rho}_{_{\bot}}\varepsilon^{\sigma},
\end{aligned}
\end{equation}
where we have defined the shorthands $$x_{+}=\frac{\omega_{1}+\omega_{2}}{m_1\omega_{2}+m_2\omega_{1}},~~ x_{-}=\frac{\omega_{1}-\omega_{2}}{m_1\omega_{2}+m_2\omega_{1}}.$$
The normalization condition is
\begin{equation}\label{mixing}
1=\int \frac{{d}^3\vec q}{(2\pi)^3} \frac{8\omega_1\omega_2{\vec{q}}^2}{3M(m_1\omega_2+m_2\omega_1)}(g_1g_2+2h_1h_2)\equiv\cos^2\theta+\sin^2\theta.
\end{equation}

In contrast to the non-relativistic wave function which only contains one independent wave function, our relativistic wave function contains four independent radial wave functions $g_1({\vec q}^2)$, $g_2({\vec q}^2)$, $h_1({\vec q}^2)$ and $h_2({\vec q}^2)$, whose numerical values are obtained by solving the instantaneous BS equation which contains four coupled equations \cite{liqiang}.
Since this method is almost complete relativistic one except the only approximation of instantaneous approach which is suitable for a heavy meson, our solutions can perfectly describe the real physical word, that is the solutions of $1^+$ states appear in pairs. For example, the first two solutions are all $1P$ states, they are $D_1$ and $D_1^{\prime}$ states with close masses, and the second two are all $2P$ states, etc, see for examples Refs. \cite{liqiang,bc}.

Using this relativistic wave function, we provide a new way to calculate the mixing angle between $^1P_1$ and $^3P_1$ states \cite{liqiang,bc}, see the second equation of Eq. (\ref{mixing}) in this paper, where the mixing angle $\theta$ is defined. Different from the usual method by using the interaction potential, the mixing angle is calculated by the relativistic wave function, which is more accurate since the wave functions is relativistic. In Eq. (\ref{wave}), $g_1({\vec q}^2)$ and $g_2({\vec q}^2)$ are $^1P_1$ waves, while $h_1({\vec q}^2)$ and $h_2({\vec q}^2)$ are $^3P_1$ waves, so the solutions of Eq.(\ref{wave}) are all mixed states of $^1P_1$ and $^3P_1$. But the right hand side of Eq. (\ref{mixing}) cannot give an unique angle with definite value and sign since it is an equation with multiple solutions. While we have given the wave functions of $^1P_1$ and $^3P_1$ states, separately in Ref. \cite{p-wave}. Comparing the numerical values of complete wave function in Eq. (\ref{wave}) for $1^+$ state with those of the $^3P_1$ wave function in Eq. (17) and $^1P_1$ wave function in Eq. (23) in Ref. \cite{p-wave}, we can uniquely determine the value and the sign of the mixing angle, and conclude that the definition of the mixing in Eq. (\ref{mixing}) is equivalent to the following commonly used formula
\begin{equation}
\begin{aligned}\label{nrmixing}
|D_1(2420)\rangle=|\frac{3}{2}\rangle=\cos\theta |^1P_1\rangle+\sin\theta|^3P_1\rangle,\\
|D_1^{\prime}(2430)\rangle=|\frac{1}{2}\rangle=-\sin\theta |^1P_1\rangle+\cos\theta|^3P_1\rangle.
\end{aligned}
\end{equation}
It should be noted that the angle here can be equivalent to $90$ degree, but it is different from the case of the equal mass system, where the meson has charge conjugate parity, the wave functions of $|^1P_1\rangle$ and $|^3P_1\rangle$ cannot exist at the same time. The equivalent $90$ degree angle here is mainly derived from Eq.(\ref{mixing}), for example, when $\theta\rightarrow -90^\circ$, in fact, it means we have the relation
\begin{equation}
\int \frac{{d}^3\vec q}{(2\pi)^3} \frac{8\omega_1\omega_2{\vec{q}}^2}{3M(m_1\omega_2+m_2\omega_1)}(g_1g_2)\rightarrow\cos^2(-90^\circ)=0.
\end{equation}
Therefore, it is not that the wave function is zero, but that the integral is zero, which is equivalent to an angle of $-90$ degree.
\subsection{Wave function of a $0^-$ state}
The relativistic wave function for the $B$ meson, which is a $0^{-}$ state, is written as \cite{Kim:2003ny},
\begin{equation}\label{pseudowave}
\begin{aligned}
\varphi_{_P}^{0^{-}}({q}_{_{\bot}})=\displaystyle
M\left(f^{}_1+f^{}_2 \frac{\not\!P}{M}-{{f_1}x_{-}\not\!{q}_{_{\bot}}}
+\frac{f^{}_2 x_{+}\not\!{q}_{_{\bot}}\not\!P}{{M}}\right)\gamma^{5},
\end{aligned}
\end{equation}
where the numerical values of radial wave functions $f^{}_1=f^{}_1({\vec q}^2)$ and $f^{}_2=f^{}_2({\vec q}^2)$ are obtained by solving the instantaneous BS equation \cite{Kim:2003ny}.
\subsection{Wave function of a $1^-$ state}
The relativistic wave function for the $D^*$ meson, which is a $1^{-}$ state, is written as \cite{1-},
\begin{equation}\begin{aligned}\label{vecwave}\varphi^{1^{-}}_{_P}({q}_{_{\bot}})&=
b_1M{\not\!\epsilon}+
b_2{\not\!\epsilon}{\not\!P}
+{b_1 M x_{-} \left(
{\epsilon}\cdot{q}_{_{\bot}}
-{\not\!{q}_{_{\bot}}}{\not\!\epsilon}\right)}
+{b_2 x_{+} ({\not\!P}{\epsilon}\cdot{q}_{_{\bot}}
-{\not\!P}{\not\!\epsilon}{\not\!{q}_{_{\bot}}})}
\\&+
{\epsilon}\cdot{q}_{_{\bot}}
\left[
\frac{b_3 {\not\!{q}_{_{\bot}}}}{M}+\frac{b_4 {\not\!P}
{\not\!{q}_{_{\bot}}}}{M^2}+\frac{(b_1 M^2+b_3 {q}_{_{\bot}}^2)x_{+}}{M}
+\frac{ (b_4 {q}_{_{\bot}}^2-b_2 M^2 )x_{-}{\not\!P}}{M^2}\right],
\end{aligned}
\end{equation}
where the four independent radial wave functions $b_1$, $b_2$, $b_3$ and $b_4$ are function of ${\vec q}^2$, and their numerical values are obtained by solving the instantaneous BS equation \cite{1-}.

\section{Transition amplitude}
In this section, we give the method of applying the relativistic wave functions to calculate the transition processes, which include the semi-leptonic decays $B\to D^{\prime}_1 \ell\nu_{\ell}$, $B\to {D}_{1} \ell\nu_{\ell}$, and strong decays ${D}^{\prime}_1 \to D^{*}\pi$, ${D}_1 \to D^{*}\pi$.
\subsection{Weak decay}
Taking the decay $B^-\to D^0_1 \ell^{-}\nu_{\ell}$ as an example, the transition amplitude can be written as
\begin{equation}\label{semi}
\mathcal{M} = \frac{G_F}{\sqrt{2}} V_{bc} \overline{u}_{\ell} \gamma_\mu (1-\gamma_5) v_{\nu_{\ell}} \left\langle D^0_1|J^{\mu}|B^- \right\rangle ,
\end{equation}
where $G_F$ is the Fermi constant, $V_{bc}$ is the CKM matrix element, $J^{\mu}=\bar{b}\gamma^{\mu}(1 - \gamma_5)c$ is the charged weak current, respectively. The hadronic matrix element in our model is expressed as the overlapping integral over the positive wave functions of the initial and final mesons \cite{Chang:2006tc}:
\begin{equation}\label{hadron}
\left\langle D^0_1(P_f,q_{_{f\bot}})|J^{\mu}|B^- (P,{q}_{_{\bot}})\right\rangle = \int \dfrac{\mathrm{d}^3 \vec{q}}{(2 \pi)^3} Tr \Big\{ \bar{\varphi}^{++}_{_{P_f}} (q_{_{f\bot}})\gamma^\mu (1-\gamma_5)  \varphi^{++}_{_P} ({q}_{_{\bot}})  \dfrac{\slashed{P}}{M}\Big\},
\end{equation}
where $q_{_{f\bot}}={q}_{_{\bot}}+\frac{m_u}{m_c+m_u}\left(P_{f}-\frac{P_f\cdot P}{M^2}P\right)$, $\bar{\varphi}^{++}_{_{P_f}} (q_{_{f\bot}})=\gamma_0\left({{\varphi}^{++}_{_{P_f}} (q_{_{f\bot}})}\right)^{\dag}\gamma_0$. $\varphi^{\pm\pm}_{_P}({q}_{_{\bot}})$ is the positive wave function, and it is defined as $\varphi^{\pm\pm}_{_P}({q}_{_{\bot}})\equiv \Lambda^{\pm}_{1}({q}_{_{\bot}})
\frac{\slashed{P}}{M}\varphi_{_P}({q}_{_{\bot}}) \frac{\slashed{P}}{M} \Lambda^{{\pm}}_{2}({q}_{_{\bot}})$, where
 the project operator
$\Lambda^{\pm}_{i}({q}_{_{\bot}})= \frac{1}{2\omega_{i}}\left[ \frac{\slashed{P}}{M}\omega_{i}\pm
(-1)^{i+1} (m_{i}+\slashed{q}_{_{\bot}})\right]$, $i=1$ and $2$ for quark and anti-quark, respectively.
\subsection{Strong decay}
Taking the OZI allowed strong decay ${D}^0_1 \to D^{*+}\pi^-$ as an example. To avoid using the wave function of light meson $\pi^-$ which may bring us large errors because the instantaneous approximation is not good for a light meson, we abandon choosing the widely used ${}^3P_0$ model, but choose the method in Ref. \cite{2632}, then the transition amplitude is written as
\begin{eqnarray}
\mathcal{M}=\frac{iP_{\pi}^{\mu}}{f_{\pi}}\langle D^{*+}|\bar d\gamma_{\mu}\gamma_{5}u|D^{0}_{1}\rangle,
\end{eqnarray}
where $P_{\pi}$ and $f_{\pi}$ are the momentum and the decay constant of $\pi^{-}$, respectively. The hadronic matrix element can be written as
\begin{equation}\label{hadron2}
\left\langle D^{*+}(P_f,q_{_{f\bot}})|\bar d\gamma_{\mu}\gamma_{5}{u}|D^0_1 (P,{q}_{_{\bot}})\right\rangle = \int \dfrac{\mathrm{d}^3 \vec{q}}{(2 \pi)^3} Tr \Big\{ \bar{\varphi}^{++}_{_{P_f}} (q_{_{f\bot}}) \dfrac{\slashed{P}}{M} \varphi^{++}_{_P} ({q}_{_{\bot}}) \gamma_\mu \gamma_5 \Big\},
\end{equation}
where $q_{_{f\bot}}={q}_{_{\bot}}-\frac{m_c}{m_c+m_d}\left(P_{f}-\frac{P_f\cdot P}{M^2}P\right)$.

\section{Results}
When solving the Salpeter equation, because the wave function is relativistic, to avoid double counting, we only need a non-relativistic interaction. So we choose the Cornell potential, a linear scalar potential plus a single gluon exchange vector potential \cite{Kim:2003ny},
$I(r)=\lambda r-\frac{4\alpha_s}{3r}+V_0$. In our calculation, we choose the following fixed parameters, $m_u=m_d=0.28$ GeV, $m_c=1.48$ GeV, $m_b=5.15$ GeV, and vary the free parameter $V_0$ to fit ground state masses. In previous paper \cite{liqiang}, we fixed others and varied the light quark mass $m_q$ to predict mixing angle, here we will vary $\lambda$ to obtain different results. Comparing with $m_q$, as a parameter of the non-perturbative linear potential, the $\lambda$ has a wide range in literature.
\subsection{Mixing angle}
When choosing our usually used value $\lambda=0.21$ GeV$^2$ \cite{gengzikan}, the branching ratio $Br(B \to D_1 \ell\nu_{\ell})$ is a little larger than data, while the $Br(B\to D^{\prime}_1 \ell\nu_{\ell})$ is much smaller than data. At the same time, we find that there is a strong cancellation between $^1P_1$ and $^3P_1$ partial waves of $D^{\prime}_1$ in the $B\to D^{\prime}_1$ transition. To see if this cancellation is sensitive to the mixing angle $\theta$, we vary the input parameter $\lambda$. A surprising phenomenon happens, the decay rate is indeed very sensitive to the mixing angle $\theta$, but not sensitive to the direct variation of parameter $\lambda$ except in some special range (note that, quark masses and $\lambda$ are our input parameters, while mixing angle is calculated by using Eq.(\ref{mixing})).

\begin{figure}
\centering
\includegraphics[width=0.43\textwidth]{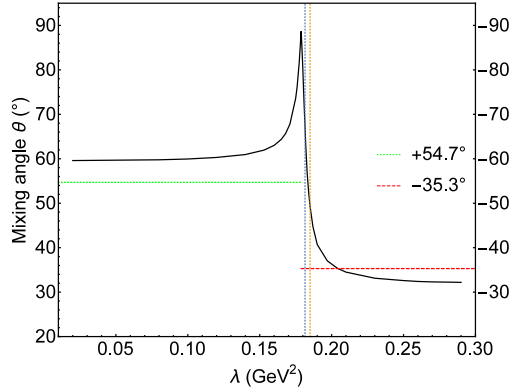}
\caption{The calculated mixing angle $\theta$ along with the parameter $\lambda$. Where $\theta >0$ with $\lambda < 0.1788~ \rm{GeV^2}$, while $\theta <0$ when $\lambda > 0.1788~ \rm{GeV^2}$. Between the blue and orange dotted lines is the range of our best results.} \label{angle}
\end{figure}

In figure \ref{angle}, for the $c\bar q$ ($q=u,~d$) $1^+$ states $D_1(2420)$ and $D_1^{\prime}(2430)$, we show the relation between the mixing angle $\theta$ and $\lambda$. When $\lambda > 0.1788~ \rm{GeV^2}$, the obtained $\theta$ is negative; when $\lambda < 0.1788~ \rm{GeV^2}$, $\theta$ is positive; $\lambda = 0.1788~ \rm{GeV^2}$ is the coincident point of $\theta=90^{\circ}$ and $-90^{\circ}$. So $\theta=90^{\circ}$ and $-90^{\circ}$ are the same point in Fig. \ref{angle}, at this point equivalent masses $M(D_1)=M(D_1^{\prime})$ are obtained. On the two sides of this point, the mass splitting and the mixing angle flip the sign, and the former is also called the mass inversion phenomenon \cite{liqiang,schnitzer,isgur2}. So instead of varying the quark mass $m_q$ \cite{liqiang}, here by changing the parameter $\lambda$, we also obtain the flip phenomenon which confirms the previous results \cite{liqiang}. Therefore we conclude that this flip phenomenon is not caused by a certain parameter, it should be the case of physics itself.

From Fig. \ref{angle}, we can see that the values of $\theta$ are stable in a large parameter range, that is, if we vary the $\lambda$ in a large range, the $\theta$ remains almost unchanged (same phenomenon occurs in Ref. \cite{liqiang}, where the varying parameter is $m_q$). For example, when $0.2~ \rm{GeV^2}\leq\lambda\leq0.29~ \rm{GeV^2}$, $\theta$ is around $-35.3^\circ \sim -32.3^\circ$; when $0.02 ~ \rm{GeV^2} \leq\lambda\leq0.16~ \rm{GeV^2}$, we have $59.6^\circ\leq\theta\leq 63.0^\circ$.
While in some special range, $\theta$ is very sensitive to $\lambda$. See Fig. \ref{angle}, in the small range of $0.16~ \rm{GeV^2}\leq\lambda\leq0.20~ \rm{GeV^2}$, $\theta$ varies from $63.0^\circ$ to $90.0^\circ$ ($-90.0^\circ$), and further from $-90.0^\circ$ to $-35.3^\circ$. The total variation  $\bigtriangleup\theta=81.7^\circ$ is very huge.

The results indicate that, when using the mixing formula Eq. (\ref{nrmixing}) where $\theta$ is an input parameter, (a) not any $\theta$ is reasonable; for example, the angle ranges $0\sim 59.6^\circ$ and $-32.3^\circ\sim -0$ are prohibited in Fig. \ref{angle}, where we use the red dashed and green dotted line to show the angles of $-35.3^\circ$ and $54.7^\circ$ in the heavy quark limit from HQET as comparison; (b) large errors may exist. The mixing angle is parameter dependent, while in Eq. (\ref{nrmixing}), the parameters in a certain range are used to estimate some values in another range, which may lead to an uncertain error. For example, if we use the wave functions obtained at $\theta=-35.3^{\circ}$ to calculate strong decays at the angle $\theta=60.0^{\circ}$ with Eq. (\ref{nrmixing}), the results are $\Gamma(D_1\to D^{*+}\pi^-)=185$ MeV and $\Gamma(D_1^{\prime}\to D^{*+}\pi^-)=11.3$ MeV. But our direct results at $\theta=60.0^{\circ}$ are $\Gamma(D_1\to D^{*+}\pi^-)=151$ MeV and $\Gamma(D_1^{\prime}\to D^{*+}\pi^-)=17.3$ MeV.
\subsection{Strong decay widths and weak production rates}
\begin{figure}
\centering
\includegraphics[width=0.43\textwidth]{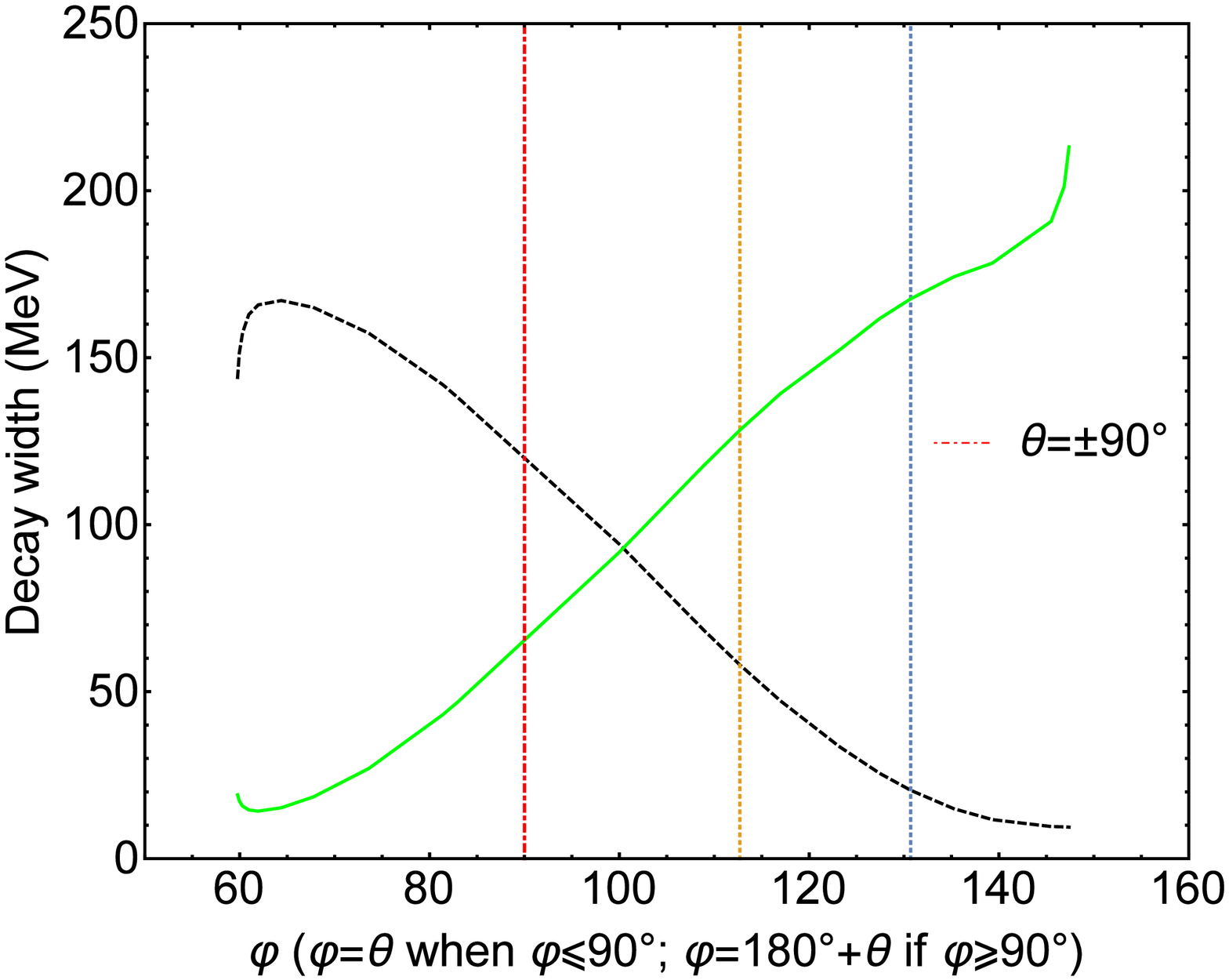}
\includegraphics[width=0.43\textwidth]{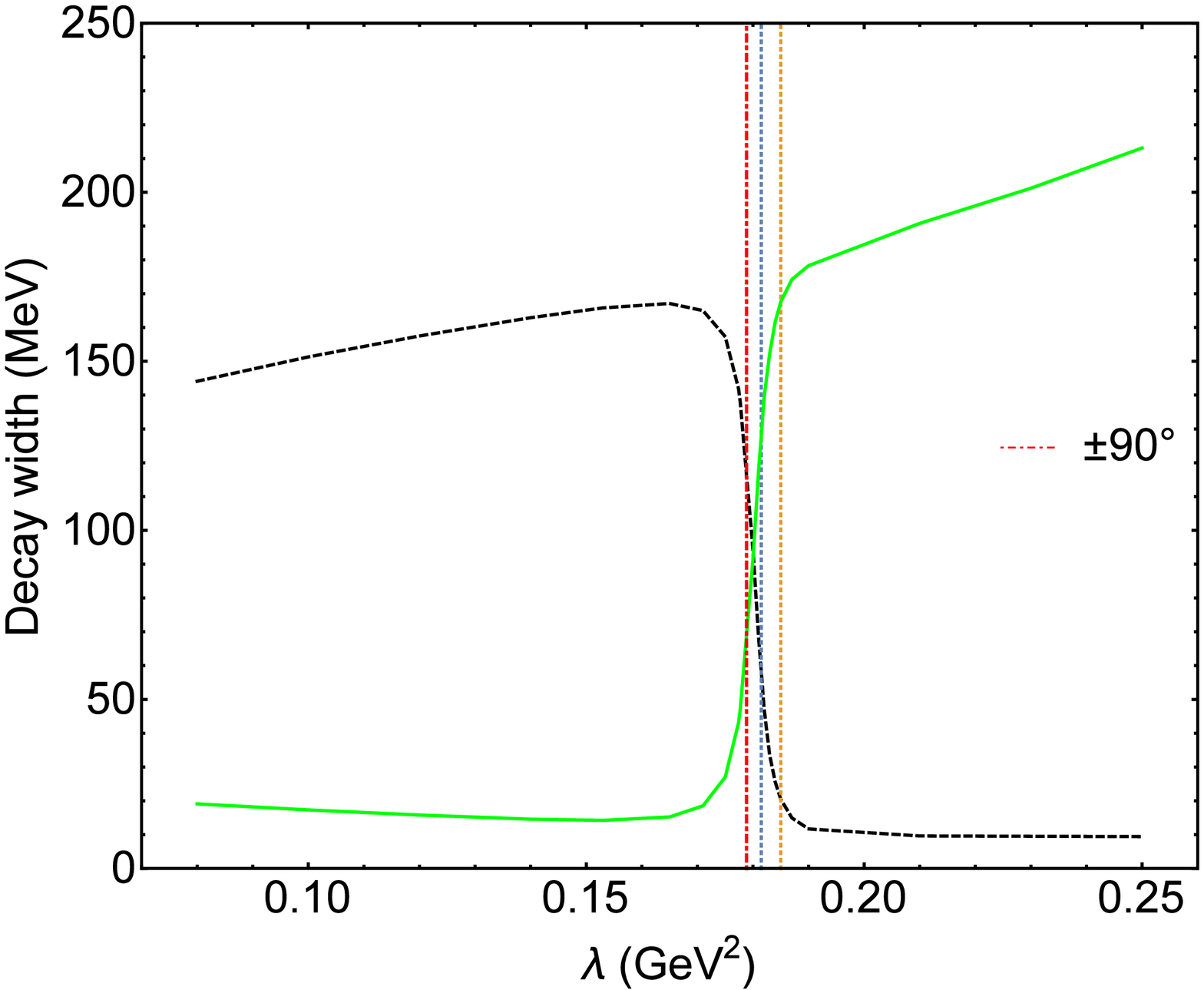}
\caption{The strong widths of the $D_1(2420)$ (black dashed line) and $D^{\prime}_1(2430)$ (green line) decay to $D^{*+}\pi^-$. Between the blue and orange dotted lines is the range of our best results.} \label{strong}
\end{figure}

\begin{figure}
\centering
\includegraphics[width=0.43\textwidth]{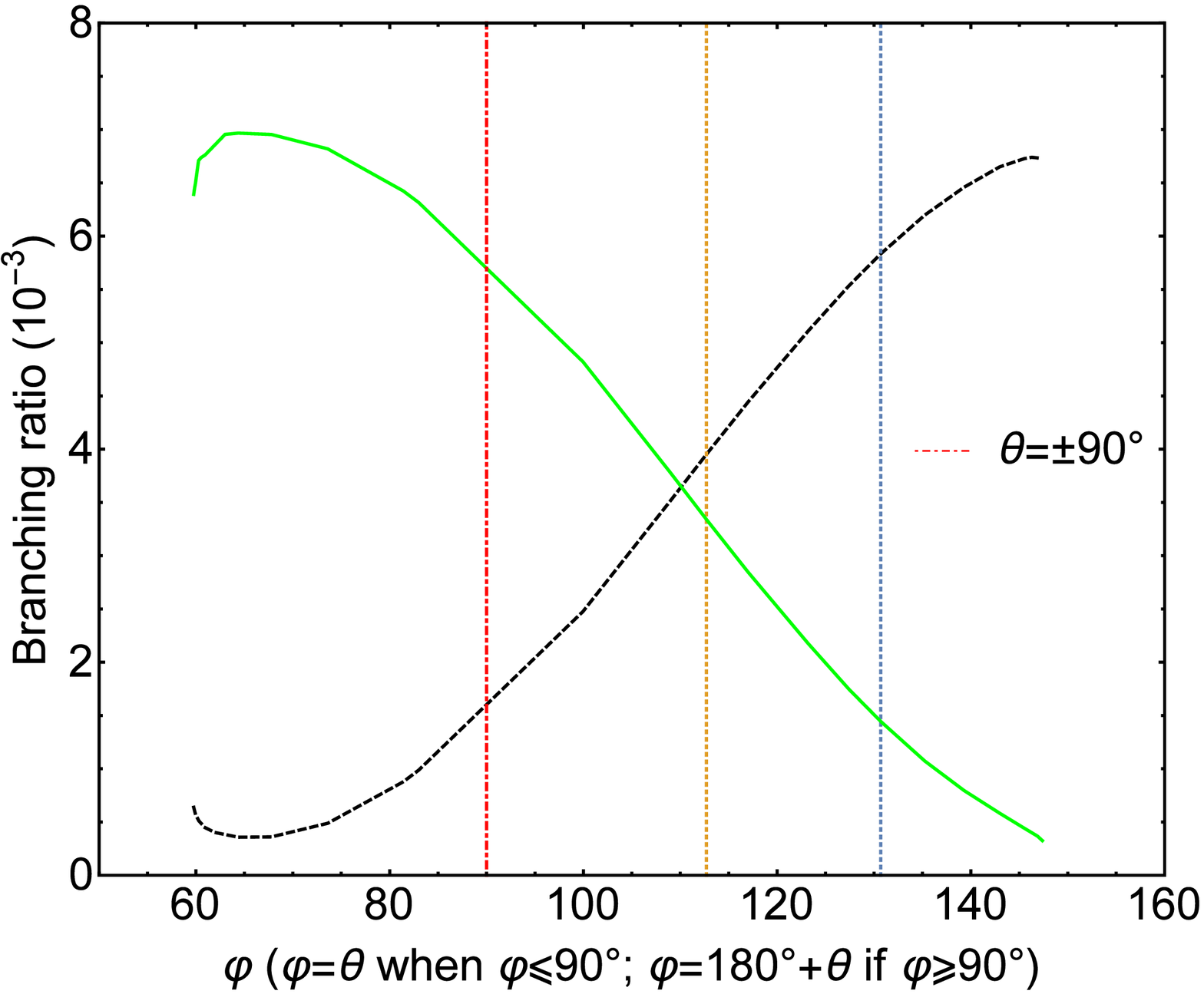}
\includegraphics[width=0.43\textwidth]{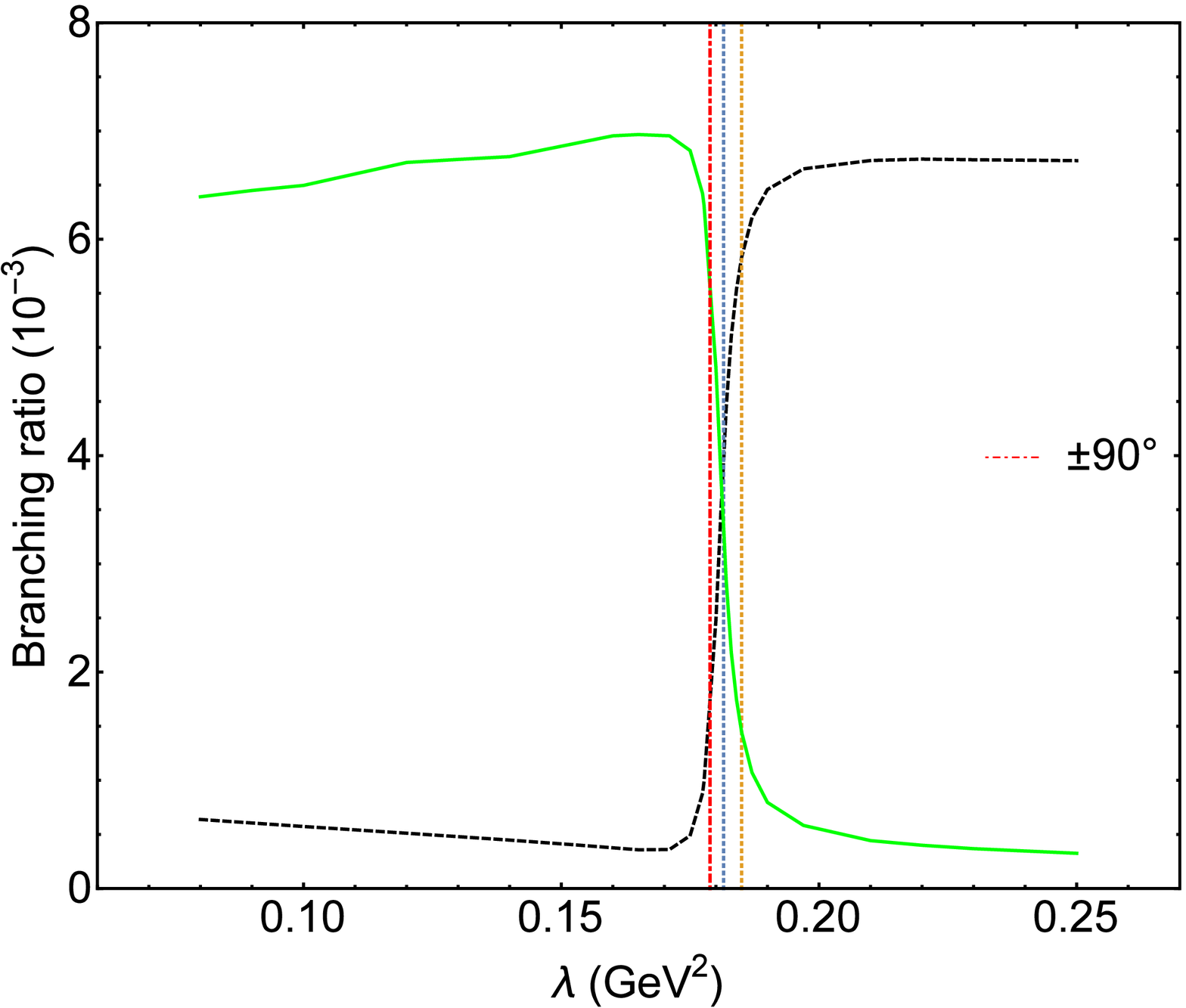}
\caption{The branching ratios $Br(B\to {D}_1\ell\nu_{\ell})$ (black dashed line) and $Br(B\to {D}_1^{\prime}\ell\nu_{\ell})$ (green line). Between the blue and orange dotted lines is the range of our best results.} \label{weak}
\end{figure}

We show the strong decay widths $\Gamma(D_1\to D^{*+}\pi^-)$ and $\Gamma(D_1^{\prime}\to D^{*+}\pi^-)$ in Fig. \ref{strong}, and the branching ratios $Br(B\to {D}_1\ell\nu_{\ell})$ and $Br(B\to {D}_1^{\prime}\ell\nu_{\ell})$ in Fig. \ref{weak}, where the black dashed line is for $D_1(2420)$ and green line is for $D^{\prime}_1(2430)$, respectively. In the left diagram in Fig. \ref{strong}, the horizontal axis is $\varphi$, which is defined as $\varphi=\theta$ when $\varphi\leq 90^{\circ}$ and $\varphi=180^{\circ}+\theta$ when $\varphi\geq 90^{\circ}$, where $\theta$ is the mixing angle. So as an example, in Fig. \ref{strong}, when $\varphi=120^{\circ}$, it is actually $\theta=-60^{\circ}$.


The strong widths and weak production rates are sensitive to the mixing angle. In most angular ranges, the curves are almost linear, see Figs. \ref{strong} and \ref{weak}. But similar to the case of mixing angle, the widths and branching ratios both remain stable (or change slightly) in a large range of parameter $\lambda$. While in a special range, they change rapidly. For example, when $0.20~ \rm{GeV^2}\leq\lambda\leq0.25~ \rm{GeV^2}$, we have $9.70~\rm{MeV} \geq \Gamma(D_1\to D^{*+}\pi^-)\geq 9.42~\rm{MeV}$, $188~\rm{MeV} \leq \Gamma(D_1^{\prime}\to D^{*+}\pi^-)\leq 213~\rm{MeV}$, but when $\lambda$ changes from $0.2~ \rm{GeV^2}$ to $0.165~ \rm{GeV^2}$, $\Gamma(D_1\to D^{*+}\pi^-)$ varies from $9.70$ MeV to $167$ MeV, $\Gamma(D_1^{\prime}\to D^{*+}\pi^-)$ from $188$ MeV to $15.2$ MeV.

%$6.71\times 10^{-3} < Br(B \to D_1\ell\nu_{\ell})< 6.73\times 10^{-3}$, and $0.325\times 10^{-3} < Br(B \to D_1\ell\nu_{\ell})< 0.494\times 10^{-3}$.
\subsection{Masses}
\begin{figure}
\centering
\includegraphics[width=0.43\textwidth]{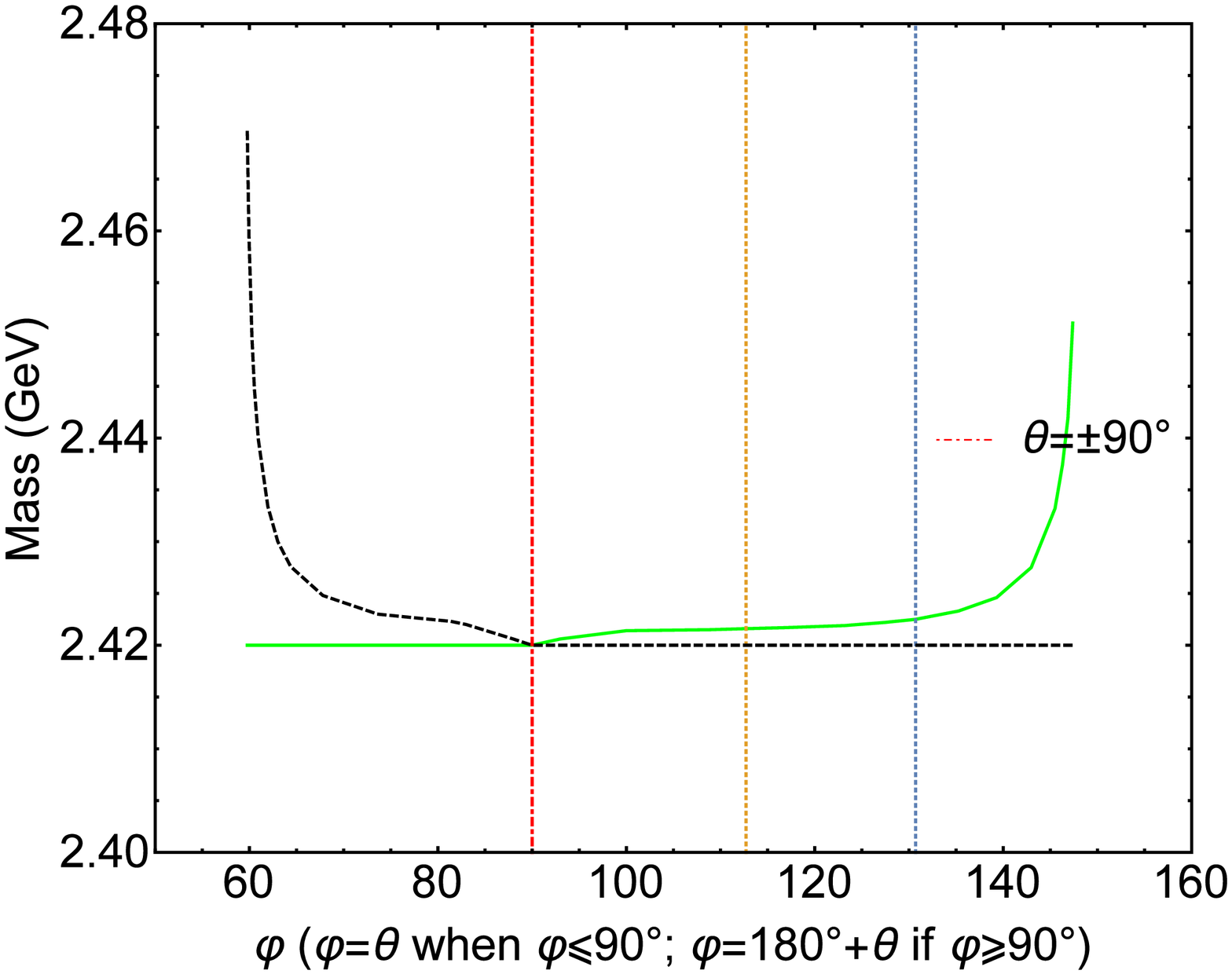}
\includegraphics[width=0.43\textwidth]{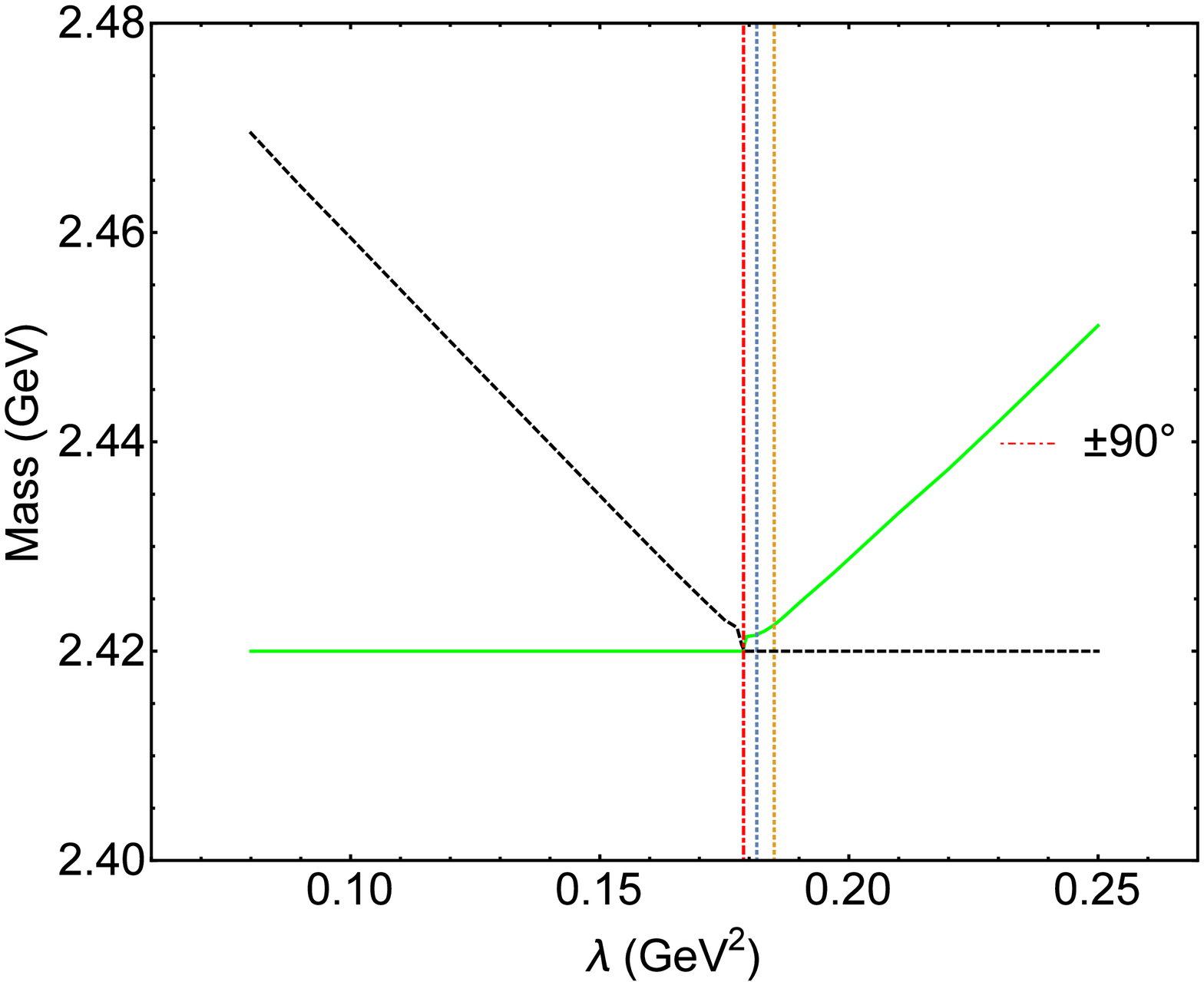}
\caption{The masses of $D_1(2420)$ (black dashed line) and $D^{\prime}_1(2430)$ (green line). Between the blue and orange dotted lines is the range of our best results.} \label{mass}
\end{figure}

The masses of $D_1(2420)$ (black dashed line) and $D^{\prime}_1(2430)$ (green line) are shown in Fig. \ref{mass}, where the lower mass $2.42$ GeV is our input, so the mass splitting $\bigtriangleup M$ between them is our prediction. The relation between the $\bigtriangleup M$ and the $\lambda$ is much different from those of strong decay width or weak production rate. The $\bigtriangleup M$ is sensitive to the $\lambda$, and the curve is perfect linear. On the contrary, the relation between the $\bigtriangleup M$ and $\theta$ is special, when the $\theta$ closes to $\pm 90^{\circ}$, the mass splitting remains small and almost unchanged. When the angle is far away from $\pm 90^{\circ}$, the $\bigtriangleup M$ changes rapidly along with the variation of $\theta$. Currently, the detected masses are \cite{PDG}
\begin{equation}
\begin{aligned}\label{m-t}
M(D_1(2420))=2422.1\pm0.8~ \rm{MeV},~
M(D^{\prime}_1(2430))=2412\pm9~ \rm{MeV}.
\end{aligned}
\end{equation}
So the experiment gives a very small mass splitting, which can be only realized with a mixing angle around the $\pm 90^{\circ}$, not the heavy quark limit $-35.3^{\circ}$ or $+54.7^{\circ}$ with a little larger splitting which conflicts with data.

We note that, the angle of mass inversion, or the condition  $M({D_1})=M({D_1^{\prime}})$,  happens at $\theta_1=\pm 90^{\circ}$, the equivalent strong decay width $\Gamma(D_1)=\Gamma(D_1^{\prime})$ happens around $\theta_2=-80.1^{\circ}$, while $Br(B\to {D}_1\ell\nu_{\ell})=Br(B\to {D}_1^{\prime}\ell\nu_{\ell})$ occurs around $\theta_3=-70^{\circ}$. These three points are not coincident, $\theta_1\neq \theta_2\neq \theta_3$, especially the last two, which enables the theoretical results and experimental data to agree with each other. Because when $Br(B\to {D}_1\ell\nu_{\ell})$ is close to $Br(B\to {D}_1^{\prime }\ell\nu_{\ell})$ at $\theta_3=-70^{\circ}$, $\Gamma(D_1)$ is still much smaller than $\Gamma(D_1^{\prime})$, this is consistent with the experimental data.

\subsection{Best fitting results}
Comprehensively considering the masses, the strong decay widths, and the weak production rates of $D_1$ and $D_1^{\prime}$, we conclude that only the mixing angle is far away from the $-35.3^{\circ}$ in HQET and moves to the direction of $-90^{\circ}$, the theory can provide consistent results with the experimental data. We find if we choose the parameter
\begin{equation}
0.1815~ \rm{GeV^2}\leq\lambda\leq0.185~ \rm{GeV^2},
\end{equation}
see the range between the blue dotted line and orange dotted line in Figs. \ref{angle}, \ref{strong}, \ref{weak} and \ref{mass}, all our predictions agree with data. \\
(1) The mixing angle
\begin{equation}
\begin{aligned}
-67.3^\circ\leq\theta\leq -49.3^\circ,
\end{aligned}
\end{equation}
which is much different from $-35.3^\circ$. This result indirectly explains why relativistic corrections or the $1/m$ corrections in HQET as well as the Lattice result do not overcome the `$1/2$ vs $3/2$' puzzle, since the corrections, whether being large or small, do not significantly change the value of mixing angle which is around $-35.3^\circ$. And the large value of mixing angle is crucial to provide consistent strong decay widths and weak production rates as well as small mass splitting of $D_1$ and $D_1^{\prime}$ with data.  \\
(2) The strong decay widths
\begin{equation}
\begin{aligned}
58.1~\rm{MeV} \geq \Gamma(D_1\to D^{*+}\pi^-)\geq 20.5 ~\rm{MeV},\\ ~128~\rm{MeV} \leq \Gamma(D_1^{\prime}\to D^{*+}\pi^-)\leq 168~\rm{MeV}.
\end{aligned}
\end{equation}
When $\lambda=0.185$ GeV$^2$, our results $\Gamma(D_1)=30.8$ MeV and $\Gamma(D_1^{\prime})=252$ MeV are consistent and close to the current data \cite{PDG}
\begin{equation}
\begin{aligned}\label{m-t}
\Gamma(D_1)=31.3\pm1.9~ \rm{MeV},\\
\Gamma(D^{\prime }_1)=314\pm29~ \rm{MeV},
\end{aligned}
\end{equation}
respectively. Where the supposition $\Gamma(D_1)=\Gamma(D_1\to D^{*}\pi)=\frac{3}{2}\times\Gamma(D_1\to D^{*+}\pi^-)$ has been used. \\
(3) The branching ratios
\begin{equation}
\begin{aligned}
 3.95\times 10^{-3} \leq Br(B \to D_1\ell\nu_{\ell})\leq 5.83\times 10^{-3},\\ ~ 3.34\times 10^{-3} \geq Br(B \to D_1^{\prime}\ell\nu_{\ell})\geq 1.44\times 10^{-3}.
\end{aligned}
\end{equation}
With $\lambda=0.1815$ GeV$^2$, the results $Br(B \to D_1\ell\nu_{\ell})\times \frac{2}{3}=2.63\times 10^{-3}$ and $Br(B \to D^{\prime}_1\ell\nu_{\ell})\times \frac{2}{3}=2.23\times 10^{-3}$ are consistent with the data
\begin{equation}
\begin{aligned}\label{brB+}
Br(B\to {D}_1\ell\nu_{\ell})\times Br({D}_1 \to D^{*+}\pi^-)=(3.03\pm 0.20)\times 10^{-3},\\
Br(B\to {D}^{\prime }_1\ell\nu_{\ell})\times Br({D}^{\prime}_1 \to D^{*+}\pi^-)=(2.7\pm 0.6)\times 10^{-3},
\end{aligned}
\end{equation}
in PDG \cite{PDG} and the average
%\begin{equation}
%\begin{aligned}\label{brB0}
%Br(B^0\to {D}_1^-\ell^+\nu_{\ell})\times Br({D}_1^{-} \to \bar{D}^{*0}\pi^-)=(2.80\pm 0.28)\times 10^{-3},\\
%Br(B^0\to {D}^{\prime -}_1\ell^+\nu_{\ell})\times Br({D}^{'-}_1 \to \bar{D}^{*0}\pi^-)=(3.1\pm 0.9)\times 10^{-3}.
%\end{aligned}
%\end{equation}
experimental results \cite{average}
\begin{equation}
\begin{aligned}\label{brB-}
Br(B\to {D}_1\ell{\nu}_{\ell})\times Br({D}_1 \to D^{*+}\pi^-)=(2.81\pm 0.25)\times 10^{-3},\\
Br(B\to {D}^{\prime}_1\ell{\nu}_{\ell})\times Br({D}^{\prime}_1 \to D^{*+}\pi^-)=(1.9\pm 0.7)\times 10^{-3}.
%Br(B^-\to {D}_1^0\ell^-\bar{\nu}_{\ell})\times Br({D}_1^0 \to D^{*+}\pi^-)=(2.81\pm 0.25)\times 10^{-3},\\
%Br(B^-\to {D}^{\prime 0}_1\ell^-\bar{\nu}_{\ell})\times Br({D}^{'0}_1 \to D^{*+}\pi^-)=(1.9\pm 0.7)\times 10^{-3},
\end{aligned}
\end{equation}
(4) The small mass splitting
\begin{equation}
\begin{aligned}\label{m-t}
1.5~\rm{MeV}\leq M(D^{\prime}_1(2430))-M(D_1(2420))\leq 2.5~ \rm{MeV},
\end{aligned}
\end{equation}
is also confirmed by data. But our result favor the mass of  $D^{\prime}_1(2430)$ being a little larger than that of $D_1(2420)$.
\section{Discussion and conclusion}
\begin{figure}
\centering
\includegraphics[width=0.43\textwidth]{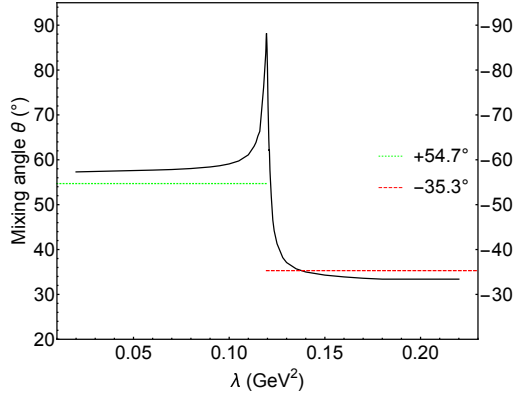}
\caption{The calculated mixing angle $\theta$ along with the parameter $\lambda$, with new setting of quark mass parameters.} \label{angle2}
\end{figure}

We have shown the relation of mixing angle $\theta$ with a varying light quark mass $m_q$ in previous paper \cite{liqiang}, which is similar to the one of `$\theta ~\rm{vs}~ \lambda$' in Fig. \ref{angle}, see the Fig. 2 in Ref. \cite{liqiang} for detail. That is, in a large $m_q$ range, the $\theta$ remains unchanged; while in some special short range of $m_q$, $\theta$ closes to $\pm90^\circ$ (where  $M({D_1})=M({D_1^{\prime}})$) and changes rapidly.
{{In Fig. \ref{angle}, we fix $m_u=0.28$ GeV, $m_c=1.48$ GeV, and show the relation of `$\theta ~\rm{vs}~ \lambda$', if we change the fixed parameters, for example, set $m_u=0.35$ GeV and $m_c=1.62$ GeV, the corresponding new relation of `$\theta ~\rm{vs}~ \lambda$' is shown in Fig. \ref{angle2}. We can see that, the large mixing angles still exists in a small special range. The obvious difference is that this special range changes and moves to the small $\lambda$ range. The curve on the left is slightly (about 2 degree) close to $+54.7^\circ$, while the right is slightly (about 1 degree) far away from $-35.3^\circ$. So we conclude that although its value depends on parameters, the existence of large angles is not caused by parameters, the physical phenomenon may be like this.}} Further, we point out that the phenomenon, that the mixing angle, strong decay widths and weak production rates remains almost unchanged over a wide range of parameters, but changes dramatically in some particular parameter range (around $\theta=\pm90^\circ$), is not unique to our model, but should exist in any method when the  relativistic corrections are considered completely.

In conclusion, using an almost complete relativistic method, we study the mixing angle, the masses and strong decays widths of $D_1$ and $D_1^{\prime}$ as well as their weak production rates in $B$ semileptonic decays. We find in a small special range of parameter $0.1815~ \rm{GeV^2}\leq\lambda\leq0.185~ \rm{GeV^2}$, theoretical results are all consistent with data and solve the puzzle of `$1/2$ vs $3/2$'.

The small mass splitting between $D_1$ and $D_1^{\prime}$, which is confirmed by experimental data, is crucial to solve the puzzle, since it means that the physics happens to be close to the mass inversion point $M({D_1})=M({D_1^{\prime}})$ ($\theta=\pm90^\circ$), where the mixing angle is large and far away from the $-35.3^{\circ}$ in heavy quark limit. And the large mixing angle leads to the solution of the `$1/2$ vs $3/2$' puzzle.

%Our results favor a little larger mass of  $D^{\prime}_1(2430)$ than that of $D_1(2420)$, otherwise, the mixing angle is positive, and we have $\Gamma(D_1)\gg\Gamma(D_1^{\prime})$ and $Br(B\to D_1 \ell\nu_{\ell})\ll Br(B\to D^{\prime}_1 \ell\nu_{\ell})$, which conflict with data.

\vspace{0.7cm} {\bf Acknowledgments}
This work was supported in part by the National Natural Science Foundation of China (NSFC) under the Grants Nos. 12075073,11865001, 12005169, 12075074, 12175025, 12075301 and 12047503, the Natural Science Foundation of Hebei province under the Grant No. A2021201009, and Natural Science Basic Research Program of Shaanxi under the Grant No. 2021JQ-074.

\end{document}